\documentclass[%
 reprint,
 amsmath,amssymb,
 aps,
]{revtex4-2}

\usepackage{graphicx}
\usepackage{dcolumn}
\usepackage{bm}

\begin{document}

\preprint{APS/123-QED}

\title{Etching Properties of PIN-PMN-PT and  \\ its Integration in Electro-Optic Phase Modulators}

\author{Salvador Poveda-Hospital}
\author{Nicol\'as Quesada }
\email{nicolas.quesada@polymtl.ca}
\author{Yves-Alain Peter } 
\email{yves-alain.peter@polymtl.ca}

\affiliation{%
 Department of Engineering Physics, \'Ecole Polytechnique de Montréal, Montréal, QC, H3T 1J4, Canada
}%


\date{\today}

\begin{abstract}
We investigate the etching characteristics of PIN-PMN-PT crystals, exploring various wet etching solutions, reactive ion etching gases, and ion beam etching. The etch rate and surface roughness are systematically measured for each method. 
A process flow to obtain PIN-PMN-PT thin films is described and performed to later build a single mode rib waveguide. An integrated electro-optic phase modulator is fabricated by depositing electrodes on the sides of the waveguides. The electro-optic coefficient implied in our phase modulators is the $r_{51}/r_{42}$. The performances are tested achieving a $V_\pi L$ of 19.2 V$\cdot$cm.
\end{abstract}

\maketitle


\section{\label{sec:level1} Introduction}

PIM-PMN-PT is a remarkable ferroelectric crystal thanks to its piezoelectric coefficient $d_{33}$ of 1830~pC/N \cite{echizenya2021pmn}. This property has caught the attention of researchers and has pushed them to develop it for applications in energy harvesting systems, ultrasound transducer, actuators, sensors, RF filters and resonators \cite{rupitsch2019piezoelectric}.
The continuous-feeding Bridgman growth technique \cite{echizenya2020pmn} is used to grow PIN-PMN-PT  single crystal in ingots, guaranteeing perfect crystallinity throughout the ingot and enabling cost-effective mass manufacturing.
Furthermore, recent studies demonstrate that this material is birefringent and transparent in the near and mid-infrared \cite{liu2020investigation}. 
High electro-optic coefficients have already been reported.
In bulk crystals, with transmission measurement, a $r_{33}$ coefficient was measured to be 72.45~pm/V \cite{liu2020investigation}. 
This first waveguide built in PIN-PMN-PT was done by titanium in-diffusion in bulk crystals \cite{zheng2023sm}, a common method due to its simplicity in inducing a local refractive index change. These titanium in-diffusion waveguides were built on Sm-doped PIN-PMN-PT bulk crystals, reporting a $r_{33}$ coefficient up to 900~pm/V. 

Developing applications in this type of crystals opens up new avenues for developing innovative technologies, by bridging piezoelectric MEMS with integrated photonics, enabling the co-integration of MEMS and photonics active components.
Despite its high potential, there is a lack of literature regarding its etching techniques. Hence the motivation to test the etch characteristics of this material with many methods. In this study we will focus on wet etching with a variety of solutions, reactive ion etching with a variety of gases and ion beam etching.

After acquiring the critical knowledge of how to precisely etch the material, we developed a process flow to obtain a thin film with a top-down approach. 
The top-down approach is not commonly used for obtaining thin films due to its complexity. Techniques like deposition offer a more straightforward alternative. However, for electro-optic effects, having a good stoichiometry and crystallinity is critical, PIN-PMN-PT, being composed of six different atoms makes its accurate deposition very challenging.
The top-down process requires having a substrate and a carrier wafer. Firstly, the single crystal substrate is bonded to the carrier wafer. Then the substrate is thinned until obtaining the desired thickness. Once the thin film is obtained, we fabricated an integrated electro-optic phase modulator by micro-fabricating waveguides and depositing electrodes on the sides. 
The integration of electro-optic materials is motivated by the fact that the phase shift is more efficient, since the refractive index will vary exactly where the mode is confined, contrary to having passive waveguides and depositing an electro-optic material on top \cite{alexander2024manufacturable}, where only part of the mode is in the electro-optic material.
We also derive the theory of the refractive index modulation for birefringent  materials, whose results are significantly different than for isotropic materials. 
The electro-optic coefficients involved in our device are  $r_{51}$ and $r_{42}$.
And we conclude by characterizing the performance of the phase modulators. We characterize two modulators where the electrodes are separated by different distances proving that the results are coherent and consistent.

\section{\label{sec:level2}Materials and Methods}

PIN-PMN-PT single crystal wafers were acquired from JFE-Mineral company in an as-cut state. 
The initial phase of processing involves planarization using chemical mechanical polishing (CMP) to attain a surface average roughness ($S_a$) below 5 nm. 
The wafer is then cleaned using standard clean 1 (SC1) at 70º, this solution is inert to PIN-PMN-PT. 
SC1 consists of one part of deionized water, one part of ammonia water and one part of hydrogen peroxide. This solution is appropriate to remove slurry that adheres tightly to the wafer during polishing,
in addition to being a standard to remove organic contaminants.
To prepare the sample for etching tests, the wafer undergoes photo-lithography patterning, followed by etching and removal of the mask. 
Lastly, the etch depth and surface roughness are quantified utilizing white light interferometry and validated through examination with an electron microscope. 
Since different etching recipes can yield to disparate etch depths, we have etched up to one micron for the solutions having a high etch rate and one hundred nanometers for the solutions having a slow etch rate.
Wet and dry etching techniques are investigated. Dry etching techniques were further categorized into two types: reactive ion etching (RIE) and ion beam etching (IBE). 

Once one side of the substrate is planarized, the substrate is then bonded to a silicon carrier wafer with a layer of Cyclotene 3022-57 used as an adhesive. Since silicon has a higher refractive index than PIN-PMN-PT, cyclotene also serves as a cladding layer since it has a refractive index of around 1.5, so that light is confined into the PIN-PMN-PT.
Cyclotene is spin coated at 2000~rpm on both the silicon carrier wafer and the PIN-PMN-PT, to promote better adhesion. 
It is then hard baked at 250~\textcelsius~for 30 minutes to achieve full polymerization and later be more resistant to wet etching and solvents. The CMP and SC1 cleaning is then repeated for the other side of the substrate.
The substrate is initially 200~µm thick, it needs to be thinned down to achieve a thin film before starting the fabrication of the device.
A mixture of 0.2\% of hydrofluoric acid (HF) and 15\% hydrochloric acid (HCl) achieves an optimum etch rate of 120~nm/min and a low roughness increase, this mixture is therefore used to thin the PIN-PMN-PT layer. Once the desired thickness is achieved the wafer is replanarized.
Then, 90~nm of silicon dioxide and 50~nm of chromium are evaporated. A lithography is performed with the negative pattern of the electrodes and waveguides, using the AZ900MIR photoresist. Wet etching with Chromium Etchant 1020 is done to remove the undesired chromium. Then, 2.5~\textmu m are etched by ion beam etching to obtain the waveguides.
The wafer is finally diced to obtain the final device, ensuring smooth edges.

\begin{figure}
    \centering
    \includegraphics[width=1\linewidth]{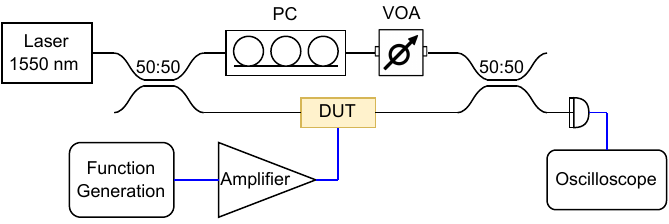}
    \caption{Mach-Zehnder interferometer schematic to characterize the phase modulator}
    \label{fig:ModC}
\end{figure}

To characterize the phase modulator a table-top Mach-Zehnder interferometer is used as shown in Fig.~\ref{fig:ModC}. 
The device is positioned in one arm, with light coupled through fiber edge-coupling. In the other arm, a polarization controller (PC) is positioned to match with the polarization of the first arm during interference, while a variable optical attenuator (VOA) is inserted to compensate for the device's losses.

\section{Rib waveguide and electro-optics theory }

\subsection{Rib waveguide design}

\begin{figure}
    \centering
    \includegraphics[width=\linewidth]{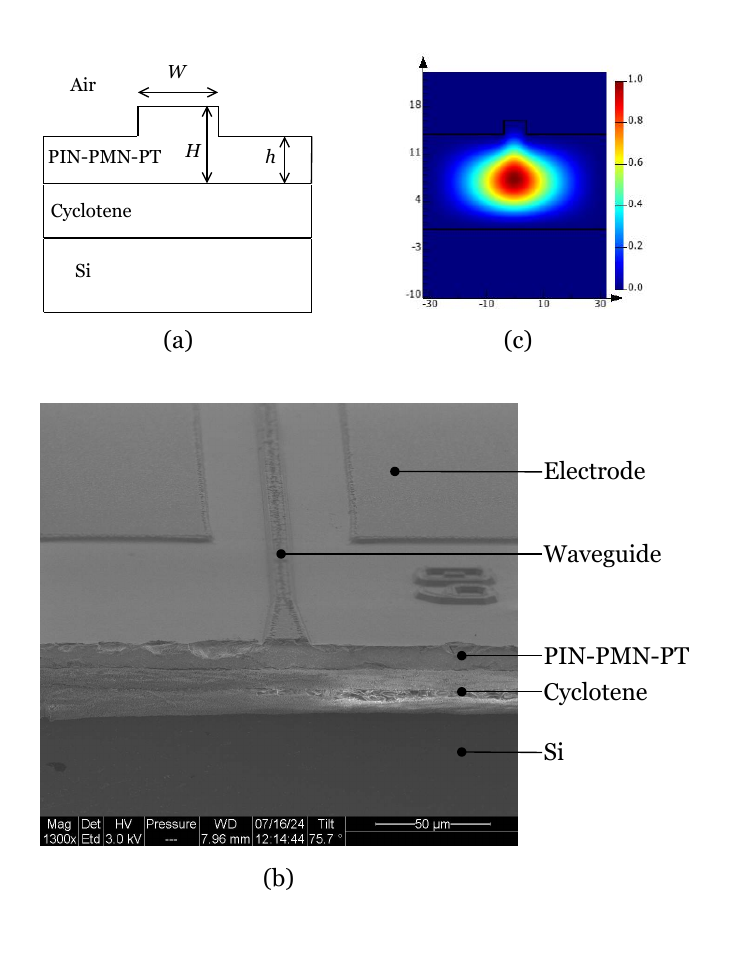}
    \caption{(a) Schematic of a rib waveguide cross-section. (b) Scanning electron microscope image of the chip edge. (c) Numerical simulation of fundamental TE mode.}
    \label{fig:RibWgSketch}
\end{figure}

For modulation and interferometric purposes, a single-mode waveguide is required to ensure that only the fundamental TE and TM modes exist, preventing any cross-modulation from being detected at the photodiode, due to higher order modes.
A formula was proposed by  Soref \textit{et al.} \cite{soref1991large} to determine the rib waveguide dimensions that will allow only the fundamental TE or TM mode to propagate
\begin{align}
\label{eq:ine1}
    \frac{h}{H} &\geq 0.5 ~, \\
     \label{eq:ine2}
    \frac{W}{H} &\leq 0.3 + \frac{h/W}{\sqrt{1-(h/H)^2}} ~,
\end{align}
where the geometric variables are sketched in Fig. \ref{fig:RibWgSketch}(a). These equations were checked experimentally \cite{rickman1994silicon, leblanc2016importance}. It is also verified numerically that only the TE modes propagate in the waveguide (Fig.~\ref{fig:RibWgSketch}(c)). 
Due to the complexity of achieving undamaged thin films under one micron using a top-down fabrication approach, micro-fabricating thick rib waveguides is the most convenient solution. The proposed dimensions are $W=5$~\textmu m, $H=12.5$~\textmu m and $h=10$~\textmu m, respecting the inequalities \eqref{eq:ine1} and \eqref{eq:ine2}.

\subsection{Electro-optic theory}

Using the methodology described above, a 12.5 µm thin film was obtained, and electrodes were evaporated with a separation distance of 80 µm and 60 µm (Fig.~\ref{fig:RibWgSketch}(b)). The cladding of the slab waveguide consists of air on the top and a 9 µm thick bottom layer of cyclotene, which provides sufficient confinement for the 1550 nm wavelength within the waveguide. With the crystal oriented in the $Z$ direction, the external electric field generated by the electrodes lies in the ($XY$) plane. For simplicity, symmetry and without of loss of generality,  we assume the field is in the $X$ direction ($E_X$), and thus we use the $r_{51}$ coefficient; equivalently, we could assume it is in the $Y$ direction and use the $r_{42}$ coefficient. The refractive index ellipsoid is then
\begin{align}
	1 =
	\frac{X^2}{n_o^2}  +
	\frac{Y^2}{n_o^2} +
	\frac{Z^2}{n_e^2} +
	2 r_{51} E_X XZ ~,
\end{align}
where $n_o$ is the ordinary refractive index in the ($XY$) plane and $n_e$ is the extraordinary refractive index in the $Z$ axis.
An orthogonal decomposition is done to obtain the eigenvalues and eigenvectors, the refractive index ellipsoid becomes
\begin{align}
\begin{aligned}
	1 &= \frac{y^2}{n_o^2} \\
	&+ \frac{n_e^2 + n_o^2 + \sqrt{(2 r_{51} E_X n_e^2 n_o^2 )^2 + (n_e^2 -n_o^2 )^2}}{2n_e^2 n_o^2 } x^2 \\
	&+ \frac{n_e^2 + n_o^2 - \sqrt{(2 r_{51} E_X n_e^2 n_o^2 )^2 + (n_e^2 -n_o^2 )^2}}{2n_e^2 n_o^2 } z^2
\end{aligned}
\end{align}
where $(x,y,z)$ are the new principal-axis coordinate. This new coordinate system correspond to the eigenvector solution
\begin{align}
    \begin{aligned}
        &x =  \\
        &\left( 
            \frac{     n_e^2 - n_o^2
            + \sqrt{ (2r_{51} E_X ne^2no^2)^2 + \left( n_e^2 - n_o^2 \right)^2} 
            }{2 r_{51} E_X ne^2no^2}
        \right)  X + Z , \\
    \end{aligned}
\end{align}
\begin{align}
    \begin{aligned}
       &z = \\
       &\left( 
            \frac{     n_e^2 - n_o^2
            - \sqrt{ (2r_{51} E_X ne^2no^2)^2 + \left( n_e^2 - n_o^2 \right)^2} 
            }{2 r_{51} E_X ne^2no^2}
        \right)  X + Z . 
    \end{aligned}
\end{align}
Since the crystal is birefringent and $2 r_{51} E_X n_e^2 n_o^2 \ll n_e^2 -n_o^2$, the binomial approximation is applied giving that the new principal-axis coordinate can be approximated to the crystal axis. The approximated ellipsoid is then
\begin{align}
\begin{aligned}
	&1 = \\
	&\left( \frac{1}{n_o^2} + 2 r_{51} E_X \right) X^2 +
	\frac{1}{n_o^2} Y^2 +
	\left( \frac{1}{n_e^2} - 2 r_{51} E_X \right) Z^2 ~.
\end{aligned}
\end{align}
The new refractive indices are given by
\begin{align}
	n_o' &= n_o - \frac{1}{2} n_o^3 r_{51} E_X ~,\\
	n_e' &= n_e + \frac{1}{2} n_e^3 r_{51} E_X ~.
\end{align}
These theoretical results are significant because, in isotropic materials, the presence of cross-terms ($YZ$, $XZ$, $XY$) results in the new principal axes being rotated by 45° relative to the crystal axes, leading to the formation of elliptically polarized waves. However, as demonstrated, in anisotropic materials, the presence of cross-terms $YZ$ and $XZ$ results in only a minimal rotation of the new principal axes relative to the crystal axes. 
Therefore, in the electro-optic modulator fabricated in this paper, the electro-optic coefficient implied is the $r_{51}/r_{42}$ and we can consider that at the output of the waveguide the light-wave will still be linearly polarized.

\section{Etching Results and Discussion}


\subsection{Wet etching}

\begin{table*}[htbp]
\caption{Experimental wet etch measurements at room temperature. Third column shows the surface roughness divided by the total etch depth. The fourth column shows the roughness increase.}
\begin{center}
\begin{tabular}{|c|c|c|c|}
\hline
\textbf{Solution}	& \textbf{Etch rate [nm/min]} 	& $\bm{S_a / d}$ \textbf{[\%]} & $\bm{Sa}$ \textbf{[nm]}  \\
\hline
BOE		& $7.2^\mathrm{b}$	& $>6^\mathrm{a}$  &  78\\
HF	1\%	& $43^\mathrm{b}$	& $>6^\mathrm{a}$  &  464 \\
HF 1.65\% + HCl 34\%		& 927	& $>6$ & 60\\
HF 0.83\% + HCl 34\%		& 436	& 4.3 & 37 \\
HF 0.49\% + HCl 29\% 		& 265	& 1.8 & 9.7\\
HF 0.2\% + HCl 15\% 		& 119	& 1.6 & 8.7\\
HCl 37\% 	& $4^\mathrm{b}$	& $<1$ & $<5$ \\
HF 0.2\% + HNO\textsubscript{3} 10\% + HCl 10\%  	& 121	& 5.1 & 51\\
HF 0.4\% + HNO\textsubscript{3} 20\% 	& 214	& 5.7 & 57 \\
HF 0.2\% + HNO\textsubscript{3} 20\% 	& 120	& 1.8 & 18 \\
Aqua Regia	& $3^\mathrm{b}$	& $<1$  & $<5$ \\
Aluminum Etchant Type A		& 0	& - & - \\
Chromium Etchant 1020 & 0 & - & - \\
SC1 (at 70º) & 0 & - & - \\
\hline
\multicolumn{3}{l}{$^{\mathrm{a}}$Redeposition is observed on the surface.} \\
\multicolumn{3}{l}{$^{\mathrm{b}}$Measured after 30 minutes. Etch rate decreases over time.} 
\end{tabular}
\label{tab1}
\end{center}
\end{table*}

Wet etching results are summarized in Table~\ref{tab1}.
Hydrochloric acid (HCl), nitric acid (HNO\textsubscript{3}) and Aqua Regia, which is a 3:1 solution of HCl and HNO\textsubscript{3}, etch very slowly, with low roughness increase. Nevertheless, due to the low etch rate, no precise roughness could be measured. 
HCl, HNO\textsubscript{3} and Aqua Regia cannot be used as etchants, at least when used individually, as their etch rate decays over time.

Hydrofluoric acid (HF) and buffered oxide etcher (BOE) etch the material; however, some reaction products do not remain in solution and are redeposited, or insoluble products are generated, precipitating on the surface.
The reaction of HF with PIN-PMN-PT produces water and various fluoride compounds such as lead fluoride, titanium fluoride, magnesium fluoride, niobium fluoride and indium fluoride. Lead fluoride and magnesium fluoride are poorly soluble in water, explaining the precipitation of insoluble products and the appearance of the white powder on the surface. Nonetheless, lead fluoride is soluble in HCl and HNO\textsubscript{3} and magnesium fluoride is soluble in HNO\textsubscript{3}. The mixture of HF, HCl and HNO\textsubscript{3} has a high etch rate, explained by the fact that, HF breaks the bounds of the initial crystal and the fluoride compounds are then soluble in water, HCl or HNO\textsubscript{3}. As observed in Table~\ref{tab1}, increasing the HF concentration linearly increases the etch rate. The relationship can be expressed as
\begin{align}
    \text{etch~rate} = 554 \times [HF] \%
\end{align}
where $[HF] \%$ is the concentration of HF in percentage and the etch rate is in nanometers per minute. Due to the reasoning explained above, only enough HCl or HNO\textsubscript{3} or a mixture of both solution are needed so that fluorides do not precipitate, consequently, increasing the concentration in HCl or HNO\textsubscript{3} does not increase the etch rate.
It has also been verified that the etch rate does not decrease over time, providing us with indications that there is no redeposition or alterations in the crystal stoichiometry that could affect the reaction.
Even though HCl or HNO\textsubscript{3} concentration do not have a linear effect on the etch rate, its selection is directly related to the roughness. 
For the HF plus HCl or HNO\textsubscript{3} mixture the roughness increases exponentially depending on the HF concentration. 
When choosing the proportion of HF, there will be a trade-off between the etch rate and roughness.

Table~\ref{tab1} also shows how the material is inert to Aluminum Etchant Type A which is a solution of phosphoric acid, nitric acid and acetic acid, and it is also inert to Chromium Etchant 1020 which is a mixture of perchloric acid, and ceric ammonium nitrate. These results allow for the fabrication of hard masks, as well as electrodes, made of aluminum or chromium.

\subsection{Dry etching}

\begin{table}[htbp]
\caption{Experimental RIE etch measurements}
\begin{center}
\begin{tabular}{|c|c|c|}
\hline
\textbf{Gas} & \textbf{Etch rate}  & $\bm{R_a / d}$ \\
             & \textbf{[nm/min]} & \textbf{[\%]} \\
\hline
Ar, 10 sccm, 300 W & 2.8 & $<1$ \\
O\textsubscript{2}/CHF\textsubscript{3}, 2/12 sccm, 200 W & 0 & -\\
SF\textsubscript{6}, 3 sccm, 200 W & 0 & - \\
\hline
\end{tabular}
\label{tab2}
\end{center}
\end{table}

As shown in Table~\ref{tab2}, sulfur hexafluoride (SF\textsubscript{6}) and the mixture of fluoroform (CHF\textsubscript{3}) and oxygen (O\textsubscript{2}) do not etch PIN-PMN-PT. Only argon (Ar) slowly etches the material, the etch process is primarily physical bombardment.

IBE etching performance is also tested, with an argon gas flow of 14~sccm, a beam voltage of 400~V,
a beam current of 80~mA,
an accelerator voltage of 60~V,
a discharge current of 1.42~A and 
a discharge voltage of 40~V, 
the etch rate is 15.6~nm/min without an increase in the surface roughness. This etch process is primarily physical bombardement.

For the RIE the resist AZ900MIR selectivity is 1:10, however for the IBE it is of 1:1, which is a significant improvement. A chromium mask for the IBE is recommended which has a selectivity of 4:1. Moreover, IBE has better directionality and much lower roughness compared to wet etching, being a necessary characteristic in some devices such as waveguides. This technique should also be employed when wanting to avoid underetching. Additionally, careful consideration should be given when using photoresist as a mask, as it may burn, as we observed when using AZ5214 photoresist.

\section{Phase modulator Results and Discussion}

\begin{figure}[ht!] 
\centering\includegraphics[width=1\linewidth]{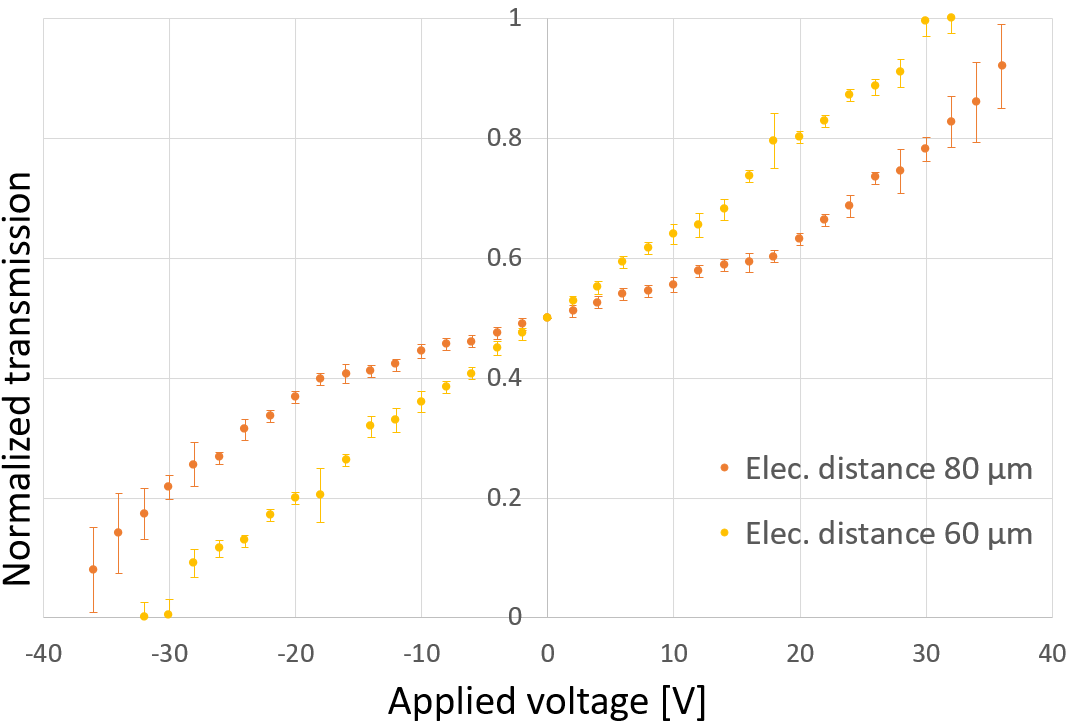}
\caption{Experimental results: normalized transmission of the MZ interferometer relative to the applied voltage.}
\label{fig:PMresutls}
\end{figure}

Two types of waveguides were microfabricated, one where the electrodes are separated by 80~\textmu m and another where they are separated by 60~\textmu m (Fig.~\ref{fig:RibWgSketch}(b)).
A sinusoidal signal at 50~kHz is generated by the function generator and an amplitude sweep from 0 to 40~V is performed. 
Above 100~V peak-to-peak the electrodes get damaged. 
The measured transmission is shown in Fig. \ref{fig:PMresutls}. 
For the electrodes separated by 60~\textmu m, the voltage needed to get a $\pi$ phase shift, $V_\pi$, is 64~V. The electrode's length is 3~mm, leading to a $V_\pi L$ of 19.2~V$\cdot$cm. 
For the electrodes separated by 80~\textmu m, the voltage needed to get a $\pi$ phase shift is 82~V, leading to a $V_\pi L$ of 24.6~V$\cdot$cm. 

The phase shift follows the formula 
 \begin{align} \label{eq:phaseR}
  	\phi =   \left( \frac{1}{2} n_o^3r_{51}  \frac{V} {d_\text{elec}}  \right)
   	\frac{2\pi L}{\lambda}
 \end{align}
 with $d_\text{elec}$ the distance between the electrodes and $L$ the length of the electrodes (the interaction length). Then, $V_\pi$ is inversely proportional to the distance between electrodes, resulting in the relationship $d_\text{elec}(80~\mu m) / d_\text{elec}(60~\mu m) = 80/60 \approx 1.33$. The experimentally measured $V_\pi$ yields a relationship of $V_\pi(80~\mu m) / V_\pi(60~\mu m) = 82/64 \approx 1.28$, consistent with the theoretical predictions.

Using a table-top Mach-Zehnder interferometer introduces significant noise. 
This noise may be attributed, in part, to the modulation of leaky modes and to the thermal drift of the fibers and the stress drift of the fibers inside the interferometer. To mitigate this noise, integration of the Mach-Zehnder interferometer is recommended. However, the proposed rib waveguide cannot be employed due to its high bending losses. 
Additionally, noise due to edge coupling could be further decreased by performing optical wire bonding \cite{luan2023towards}.
The piezoelectric properties are also a source of noise, an external voltage applied to the material yields to a mechanical deformation, which may result in minimal variations in the effective refractive index.
Furthermore, the thickness of the PIN-PMN-PT thin film should be further reduced to approximately one micron before proceeding with device fabrication. Obtaining an undamaged thin film with optimal planarity poses considerable challenges with the current top-down fabrication approach.
Decreasing the noise by integrating the device will also allow to precisely study the contribution of the linear electro-optic effect (Pockels effect) and the contribution of the quadratic electro-optic effect (Kerr effect) as already performed in similar materials such as lanthanum-modified lead zirconate titanate (PLZT) \cite{huang2020large}. 

\section{Conclusion}

Etching is a critical step for the fabrication of MEMS and photonics devices. By precisely understanding its etch characteristics, fabrication process flows can be developed. We have found that a mixture of HF and HCl/HNO\textsubscript{3} is the most appropriate for wet etching PIN-PMN-PT. This mixture has a linear etch rate of 554 nm/min
per percentage concentration of HF in the total volume. HCl/HNO\textsubscript{3} concentration does not affect the etch rate, however concentrations higher than 15\% are recommended for a correct dissolution of by-products. 
Additionally, we have also shown that PIN-PMN-PT is compatible with many metal etchants such as Aqua Regia, Aluminum Etchant Type A and Chromium Etchant 1020.
In cases where a side wall angle close to 90º is required, it is recommended to use IBE because its etching is anisotropic, although its etch rate is low.
We believe that these results will facilitate the fabrication of high quality MEMS and photonic devices.

Moreover, we successfully fabricated a PIN-PMN-PT rib waveguide. We built an electro-optic phase modulator and characterized it using a table-top Mach-Zehnder interferometer. Using the refractive index ellipsoid, we derived the refractive index change and principal axis direction, proving that for birefringent crystals the principal axis rotation with respect to the crystal axis is negligible. For the waveguide with electrodes separated by 60~\textmu m we found a $V_\pi L$ of 19.2~V$\cdot$cm and for the waveguide with electrodes separated by 80~\textmu m we found a $V_\pi L$ of 24.6~V$\cdot$cm, at a wavelength of 1550 nm.

\begin{acknowledgments}
The authors acknowledge support from the Ministère de l’Économie et de l’Innovation du Québec, the Natural Sciences and Engineering Research Council of Canada and Photonique Quantique Québec.
YA Peter thanks the Laboratoire de Microfabrication for helpful suggestions and discussions.
\end{acknowledgments}

\appendix

\nocite{*}

\bibliography{apssamp}

\end{document}